""""""""""""""""""""""""""""""""""""""""""""""""""""""""""""
\documentstyle{article}
 
\font\tenrm=cmr10 
\font\tenit=cmti10 
\font\elevenbf=cmbx10 scaled\magstep 1
\font\elevenrm=cmr10 scaled\magstep 1
\font\elevenit=cmti10 scaled\magstep 1

\font\ninerm=cmr9

\renewenvironment{thebibliography}[1]
{\elevenrm
\begin{list}{\arabic{enumi}.}
{\usecounter{enumi} \setlength{\parsep}{0pt}
\setlength{\itemsep}{3pt} \settowidth{\labelwidth}{#1.}
\sloppy
}}{\end{list}}

\newcounter{currentnumber}

\begin{document}
\begin{center}
\vglue 0.6cm
{
{\elevenbf  \vglue 10pt
TQFT versus RCFT: 3-d topological invariants\marginpar{\ninerm
\vspace{-1.5in}\hspace{-1in}hep-th/9407016} 
\marginpar{\ninerm\vspace{-1.3in}\hspace{-1in}KFT U\L\ 1/94}
\\}
\vglue 1cm
{\tenrm BOGUS\L AW BRODA\footnote{
\ninerm\baselineskip=11pt 
e-mail: {\tt bobroda@mvii.uni.lodz.pl} }
\\}
\baselineskip 13pt
{\tenit Department of Theoretical Physics, University of \L\'od\'z\\}
\baselineskip 12pt
{\tenit  Pomorska 149/153, PL--90-236 \L\'od\'z, Poland\\}}
\vglue 0.8cm
{\tenrm ABSTRACT}
\end{center}
\vglue 0.3cm
{\rightskip=3pc
\leftskip=3pc
\tenrm\baselineskip=12pt
\noindent

A straightforward relationship between the two approaches
to 3-dimensional topological invariants, one of them put
forward by Witten in the framework of topological quantum
field theory, and the second one proposed by Kohno in terms
of rational conformal field theory, is established.

}

\vglue 0.6cm
{\elevenbf\noindent 1. Introduction}
\setcounter{currentnumber}{1}
\vglue 0.4cm

\baselineskip=24pt
\elevenrm

In 1989, Witten proposed a new topological 
invariant of 3-dimensional manifolds $\tau_k({\cal M})$
in the framework of
topological quantum field theory (TQFT) defined by the
Chern-Simons action.$^1$ His idea has been next developed
and mathematically refined by Reshetikhin and Turaev.$^2$
Though, strictly speaking, Witten explicitly used the
apparatus of TQFT (or even conformal field theory) and
described 3-manifolds in terms of the Dehn (or rational)
surgery, whereas Reshetikhin and Turaev applied the
combinatorics of representations of quantum groups and the
honest (or integer) surgery, we will identify the both
methods as the Reshetikhin-Turaev-Witten (or surgical)
approach. Other topological invariant of 3-manifolds,
proposed by Turaev and Viro, bases on a simplicial
description of 3-manifolds and is related, from physical
point of view, to 3-dimensional topological
gravity (see, e.g. Ref.~3).
Finally, Kohno has introduced another invariant of
3-manifolds $\phi_k({\cal M})$, 
which relies on the formalism of rational
conformal field theory (RCFT) and the Heegaard
decomposition$^{4,5}$ (see also references in Ref.~5).

Thus, we have at our disposal the three kinds of 3-manifold
invariants, each one associated with a different
topological presentation of a 3-manifold 
and with different physical and/or
mathematical tools:
\begin{description}
\item[(i)] surgery---quantum Chern-Simons field theory (Witten),
representations  of quantum groups (Reshetikhin and Turaev); 
\item[(ii)] triangulation---representations of quantum 
groups (Turaev and Viro),
topological gravity;
\item[(iii)] Heegaard decomposition---WZW model (Kohno),
representations of mapping class group.
\end{description}
Interestingly, it appears that all the approaches are,
in fact,
equivalent. The well-known  equivalence of the
surgical and the simplicial invariant has been established
in physical context in Ref.~3, whereas the equivalence of the
surgical and the Kohno invariant in the context of quantum
field theory
(see
Theorem 4.2 in Ref.~5) is addressed in
this letter.

\vglue 0.6cm
{\elevenbf\noindent 2. 3-d invariants}
\vglue 0.4cm
In this section, we shortly summarize the definition of
the both 3-dimensional invariants,
$\tau_k({\cal M})$ and $\phi_k({\cal M})$ respectively.
\vglue 0.2cm
{\elevenit\noindent a) tqft}
\vglue 0.1cm

The surgical presentation of a  3-manifold let us associate,
highly non-uniquely (i.e.\ modulo the Kirby moves), 
to each manifold ${\cal M}$ a link ${\cal
L}$. Therefore, from the point of view of ${\cal M}$,
we should identify all the links related by
means of the two Kirby moves, as they yield the same
3-manifold. To construct the topological 
invariant of $\cal M$ we can introduce
a special element of the fusion algebra related to ${\rm SU}(2)$
Chern-Simons field theory
\begin{equation}
\Omega=\bigoplus^{k-1}_{i=0}d_iW_i{\rm,}
\end{equation}
where $k$ is a fixed positive integer (the level), $k=1,2,\ldots$,
$W_i$ is an element of the fusion algebra corresponding to
$i+1$ dimensional irreducible representation of ${\rm
SU}(2)$, and $d_i$ is a number, the quantum dimension of $W_i$.
Thus, in the context of ${\rm SU}(2)$ Chern-Simons field 
theory, $\Omega$ 
gives rise to a linear combination of
Wilson lines weighted by $d_i$.

The ``partition function''
\begin{equation}
\langle\Omega_{\cal L}\rangle{\rm,}
\end{equation}
with $\Omega$ assigned to each component of ${\cal L}$, is
invariant with respect to the second Kirby move, handle
slides. For further convenience,  we shall slightly 
change the normalization of
$\Omega$ defining 
\begin{equation}
\omega\equiv\langle\Omega_{\cal
U}\rangle^{-\frac{1}{2}}\Omega{\rm,}
\end{equation}
where ${\cal U}$ is a 0-framed unknot.
After a simple re-normalization (stabilization) we obtain a
true topological invariant 
\begin{equation}
\tau_k({\cal M})=\frac{\langle\omega_{\cal L}\rangle}{
\langle\omega_{{\cal U}_+}\rangle^{b_+}\,\langle\omega
_{{\cal U}_-}\rangle^{b_-}}{\rm,}
\end{equation}
where ${\cal U}_+$ (${\cal U}_-$) denotes an unknot with a
single positive (negative) twist, and $b_+$ ($b_-$) is the
number of positive (negative) eigenvalues of the linking
matrix $\ell k({\cal L})$.
\vglue 0.2cm
{\elevenit\noindent b) rcft}
\vglue 0.1cm
An arbitrary 3-manifold ${\cal M}$ can also be presented via
a Heegaard decomposition 
\begin{displaymath}
{\cal M}={\cal V}_g\cup_h(-{\cal
V}_g) {\rm,}
\end{displaymath}
where ${\cal V}_g$ is a handlebody of genus $g$, and
$h$ is a homeomorphism identifying the boundaries of the
handlebodies. Let denote by $\rho_k$ the projective
representation of the mapping class group acting on colored
edges of the dual graph representing a pant decomposition
of $\Sigma_g=\partial{\cal V}_g$,
found in the
context of ${\rm SU}(2)$ RCFT at the level $k$ in Ref.~6 (see
also Ref.~7). Now we can introduce a ``vacuum expectation value''
\begin{equation}
\langle\rho_k(h)\rangle_0{\rm,}
\end{equation}
where the symbol ``$\langle\;\rangle_0$'' means 
the $(0,0)$-entry of the matrix 
$\rho_k(h)$. The quantity (5) is independent of the
Heegaard decomposition modulo $\Gamma_k$, a cyclic group
generated by $\kappa_k=\exp(2\pi i \frac{c}{24})$, with $c=3k/(k+2)$,
$k=1,2,\ldots$.
Finally, following Ref.~4, we have to stabilize (5)
yielding 
\begin{equation}
\phi_k({\cal M})=\langle\omega_{\cal U}\rangle^g\;
\langle\rho_k(h)\rangle_0{\rm,}
\end{equation}
a true invariant of ${\cal M}$ (modulo $\Gamma_k$).

\vglue 0.6cm
{\elevenbf\noindent 3. Equivalence}
\vglue 0.4cm

In this section, we will show the equivalence of the (TQFT)
Chern-Simons surgical  invariant $\tau_k({\cal M})$ (Eq.~4) and the
(RCFT) Heegaard decomposition invariant
$\phi_k({\cal M})$ (Eq.~6). To this end we will translate 
$\phi_k$ (basically built
of a product of generators of Dehn twists) onto weighted
linear combinations of Wilson loops (surgery loops). 
As it is well-known, every
orientation-preserving homeomorphism of a closed orientable
2-manifold $\Sigma_g$ of genus $g$ is isotopic to a product
of Dehn twist homeomorphisms along the $3g-1$ (Lickorish) canonical curves
pictured in Fig.~1. An explicit set of generators of the
Dehn twists satisfying Wajnryb's relations for genus $g$,
modulo $\Gamma_k$, is proposed in Ref.~6. In turn, each Dehn twist
along a circle $\cal U$ can be 
implemented at the representation
level by assigning to it $\omega_{{\cal U}_\pm}$.
Heuristically, it follows from the fact, well-known in the
Fenn-Rourke version of the Kirby calculus, that
$\omega_{{\cal U}_+}\;(\omega_{{\cal U}_-})$
twists clockwise  (anti-clockwise)
lines going through ${\cal U}_+\;({\cal U}_-)$. It has
been shown in Ref.~7 explicitly that putting $\omega_{{\cal U}_\pm}$
round an edge of the dual graph representing a pant
decomposition of $\Sigma_g$ yields the corresponding Dehn
twist (modulo $\Gamma_k$). Thus, we can assign to every
product of generators of Dehn twists
$T=\tau_r\cdots\tau_2\tau_1$ a special link denoted as ${\cal
L}^{-1}(T)$. The link  consists of a finite number of
circles, each of which lies on, and is concentric with, one
of the annuli of
Fig.~2.
The arrangement of the
components of ${\cal L}^{-1}(T)$ follows from the order of
the Dehn twists $\tau_n,\;n=1,2,\ldots,r$, furthest Dehn
twists (greatest indices) correspond to outermost surgery unknots ${\cal
U}_\pm$. 
Proceeding this way we obtain the special link
${\cal L}^{-1}(T)$
corresponding to $\rho_k(h)$. To calculate the ``vacuum
expectation value'' of $\rho_k(h)$, i.e.\ the
$(0,0)$-entry (0 means the trivial representation), 
one should evaluate the ``partition
function'' of an auxiliary link ${\cal U}{\cal L}^{-1}(T)$,
derived from ${\cal L}^{-1}(T)$ in the following manner.
Since $\rho_k(h)$ is to act on a dual graph (Fig.~2) colored with
the trivial representation (the first 0-entry), actually the
graph vanishes due to the triviality of representations. To
project out the second 0-entry we should put a 0-framed
surgery unknot ${\cal U}$, i.e.\ $\omega_{\cal U}$, round each
one-handle of $\Sigma_g$ encircling lines
belonging to ${\cal L}^{-1}(T)$. $\omega_{\cal U}$ has a
well-known property of cancelling all lines going through
${\cal U}$ except those which correspond to the trivial
representation. We can note that actually the required (minimal)
number of ${\cal U}$'s (primary ones)
is exactly equal to $g$, because
temporarily we can always add some auxiliary unknot ${\cal
U}$, and use the second Kirby move to shift it on $\Sigma_g$. 
More precisely,
handle-sliding ${\cal U}$ along those (non-homotopic,
meridian-like)
${\cal U}$'s that
already reside on the handlebody we can transport it onto
any $1$-handle of $\Sigma_g$. Cutting the lines going
through ${\cal U}$ we can free ${\cal U}$ and next remove
it. Thus ${\cal U}{\cal L}^{-1}(T)$ consists of ${\cal
L}^{-1}(T)$ and $g$ 0-framed unknots (see Fig.~3). 
Then we obtain the
following equality
\begin{equation}
\langle\rho_k(h)\rangle_0=\langle\omega_{\cal
U}\rangle^{-g}\langle\omega_{{\cal U}{\cal
L}^{-1}(T)}\rangle{\rm,}
\end{equation}
which expresses $(0,0)$-entry of $\rho_k(h)$, by means of a
link invariant.  The first term on RHS of (7) assures the
right normalization of the $g$  ${\cal U}$'s. 
RHS of (7) resembles Eq.~2, but there are some important differences.
First of all, to repeat the Dehn twist construction of
${\cal M}$ in the surgery language, one is forced to put all
the canonical surgery unknots ${\cal U}_\pm$ in the
reversed order, i.e.\  according to Ref.~8,
one should excavate tunnels in the
handlebody ${\cal V}_g$ slightly 
deeper in each step.
Fortunately,
it appears that utilizing untwisted unknots ${\cal U}$ we
can reverse the order of the canonical unknots ${\cal U}_\pm$.
The method
is as follows. Temporarily adding auxiliary ${\cal U}$'s (as
it was explained earlier) we can cut out of the handlebody
the two kinds of segments and corresponding graphs $G_2$
and $G_4$ (see Fig.~4). Turning the ``ribbons'' $180^\circ$ as
shown in Fig.~4 we effectively change the order of lines.
Gluing the lines back and throwing out auxiliary ${\cal
U}$'s we obtain ${\cal UL}(T)$, a link with the
reversed order of all components (obviously, except ${\cal U}$'s).
Accordingly,
\begin{equation}
\langle\omega_{{\cal U}{\cal
L}^{-1}(T)}\rangle=\langle\omega_{{\cal U}{\cal
L}(T)}\rangle{\rm.} 
\end{equation}
Gluing back the complementary handlebody ${\cal V}_g$ 
we obtain ${\cal
M}$. Thus, by virtue of (6), (7) and (8),
\begin{equation}
\phi_k({\cal M})=\langle\omega_{{\cal UL}(T)}\rangle{\rm.}
\end{equation}
Since ${\cal UL}(T)$ provides us with a proper surgical
description of ${\cal M}$, we should only account for the
absence of the denominator present in Eq.~4. But
\begin{equation}
\langle\omega_{{\cal U}_\pm}\rangle=\kappa_k^{\mp(2k+1)}{\rm,}
\end{equation}
and therefore the expressions (4) and (9) are equivalent
modulo
$\kappa_k^{(2k+1)\sigma}$  
(where $\sigma$ is the signature of the linking matrix
$\ell k({\cal L})$), with ${\cal L}=
{\cal UL}(T)$.

In this way, we have shown that 
\begin{equation}
\tau_k({\cal M})=\phi_k({\cal M})\bmod \Gamma_k{\rm,}
\end{equation}
for an arbitrary closed oriented 3-manifold $\cal M$.
\vglue 0.6cm
{\elevenbf\noindent Acknowledgments}
\vglue 0.4cm
The work has been partially supported by the KBN grants 
2P30213906 and 2P30221706p01. When the work was in draft,
the author received several papers independently addressing
the same problem (see Ref.~9 and references therein).
\vglue 0.6cm
{\elevenbf\noindent References}
\vglue 0.4cm

\pagebreak
{\elevenbf\noindent Figure captions}
\vglue 0.4cm
\begin{description}
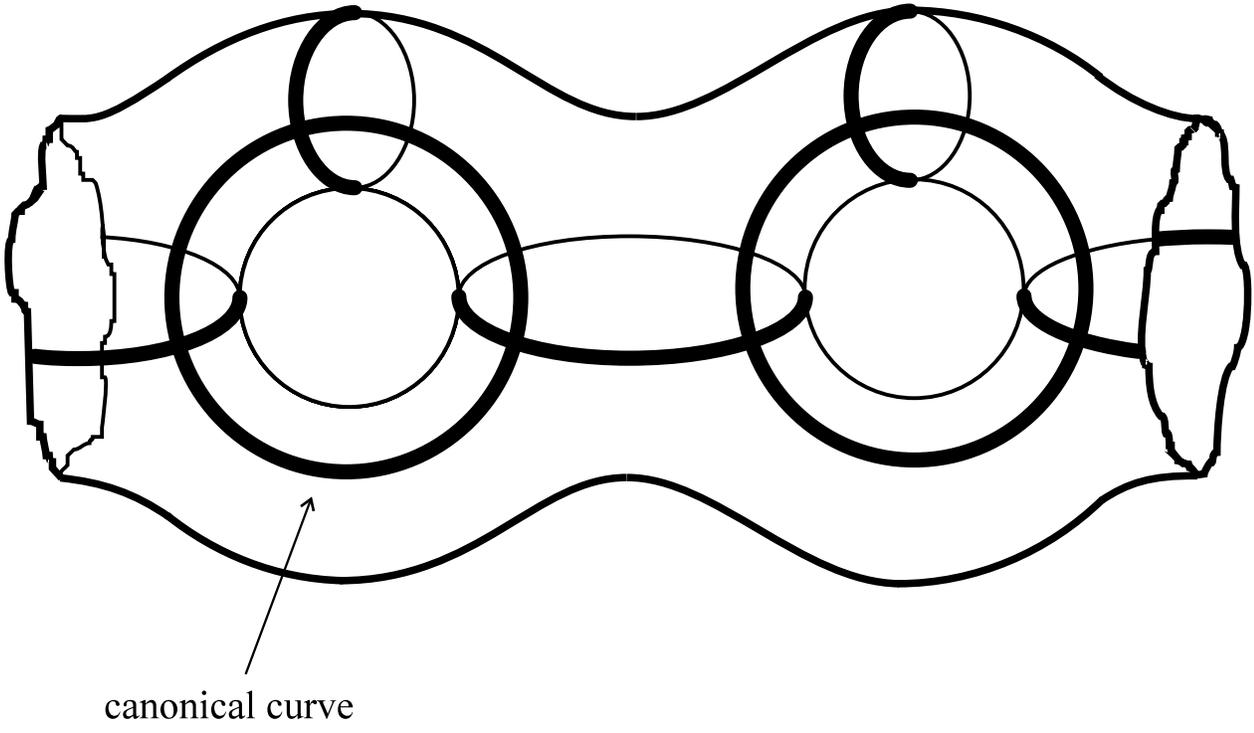
\item[Fig.~1] $3g-1$ canonical curves on a closed
orientable 2-manifold $\Sigma_g$ of genus $g$.
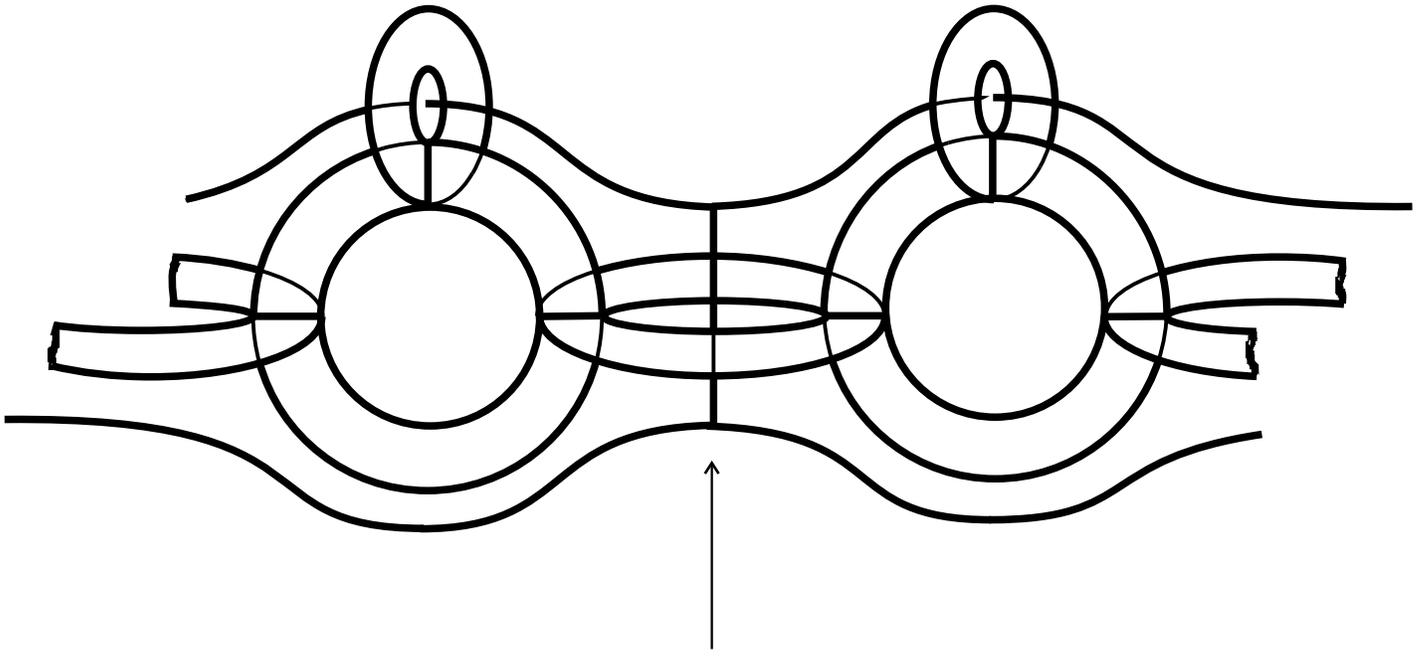
\item[Fig.~2] The annuli accommodating the circles
corresponding to the Dehn twists and a dual graph
representing a pant decomposition of $\Sigma_g$.
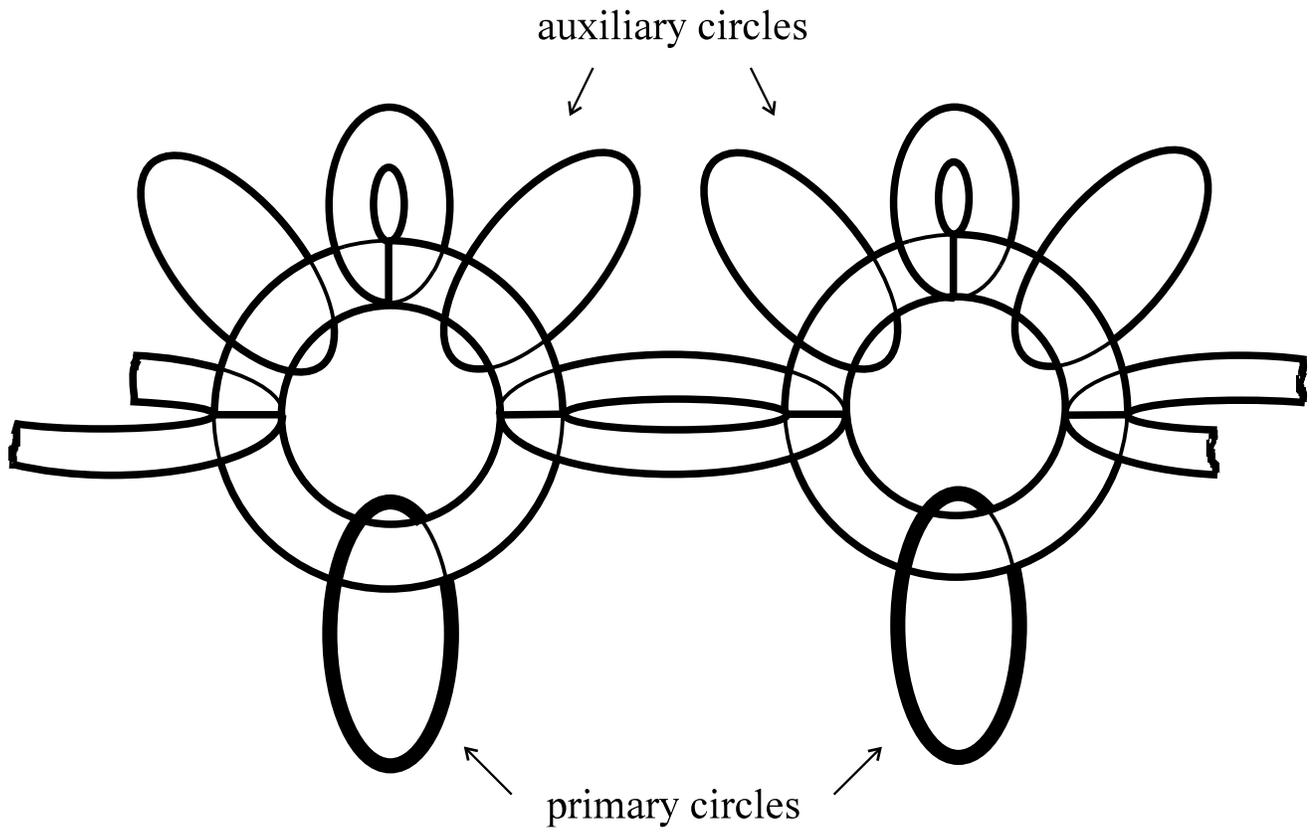
\item[Fig.~3] ${\cal UL}^{-1}(T)$ consisting of ${\cal
L}^{-1}(T)$ and $g$ primary 0-framed unknots shown in the
lower part of Figure. There are also auxiliary 0-framed
unknots in the upper part of Figure.
\item[Fig.~4a] $G_2$ cut out with 2 auxiliary 0-framed
unknots. 
\item[Fig.~4b] $G_4$ cut out with 2 auxiliary and 2 primary
0-framed unknots.
\end{description}

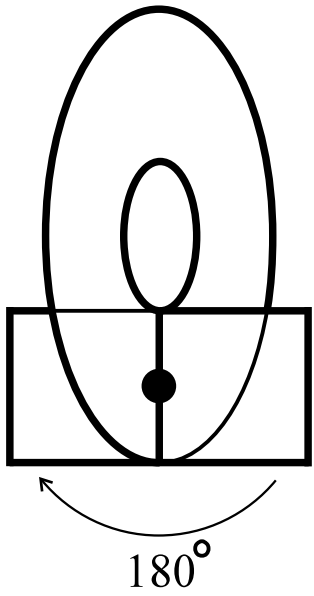
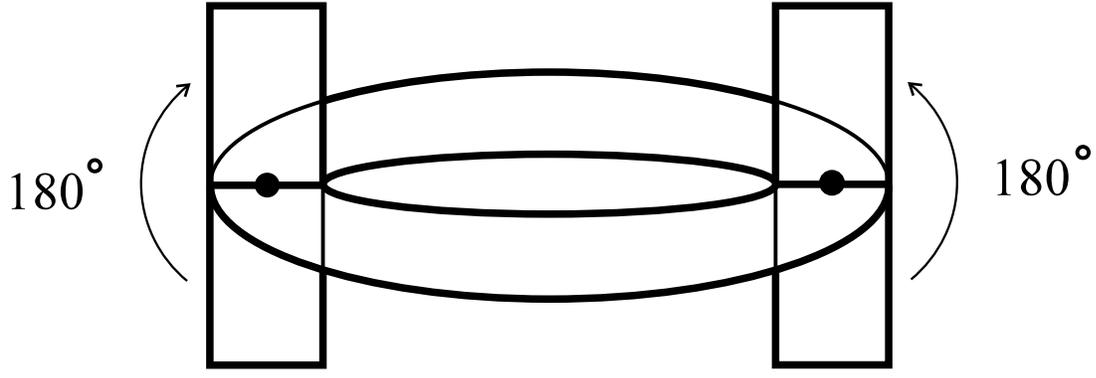
\end{document}